# Separation of particles and bio-samples by tilted-angle standing surface acoustic waves: a theoretical analysis

Wei Wang, Antoine Riaud *

*ASIC and System State Key Laboratory, School of Microelectronics, Fudan University, Shanghai 200433, China*
*Corresponding author. Email: antoine_riaud@fudan.edu.cn

**Abstract**

Separation of particles and bio-samples by the standing surface acoustic wave (SSAW) has attracted considerable interest in many biological protocols. In this work, we thoroughly analyze the behavior of particles in an SSAW field that is introduced at a random angle to the flow direction, numerically and analytically. An explicit transition threshold is deduced by the acoustic to drag force ratio between the drift mode and the locked mode, while the trajectories of particles in both of these modes are demonstrated in an analytical way. The deviation in the drift mode is revealed fundamentally nonlinear, whose slope is proportional to the sixth power of the particle radius, the fourth power of the acoustic pressure, the square of the acoustic contrast factor, and reciprocal of the square of the flow velocity. Our analytical formula shows a good agreement with numerical calculations and experimental data even in the transition regime. It provides a basis for understanding how particles and bio-samples behave in an SSAW field, and an approach to effectively evaluate the design of the acoustophoretic system.

## 1. Introduction

Separation of particles and cells is a critical step prior to chemical and biological sample analysis. Typically, it is enabled in micro-systems by a range of passive [1-7] and active methods relying on the physical characteristics of the particles. Prominent methods are inertial hydrodynamics [8-10], magnetic-labeling [11, 12], optical trapping [13], dielectrophoresis [14-16], acoustic sorting (acoustophoresis) [17, 18], and electrophoresis [19-21]. Among these techniques, acoustophoresis stands out as a label-free, contactless, non-invasive, and biocompatible manner regardless of the optical, magnetic and electrical properties of the particles and media [17, 22-30].

Particles subjected to an acoustic field will experience a steady acoustic radiation pressure. Especially, a standing acoustic field pulls the particles towards the pressure nodes [17, 31, 32] or antinodes [17, 18, 31, 33-36] depending on their acoustic contrast factors. However, the separation performance is limited by its short separation distance (normally a quarter wavelength) when the fluid flow direction is parallel to the pressure nodal lines.

To break through the limitation, a standing surface acoustic wave (SSAW) was introduced at a specific angle to the flow direction [25, 37-43]. The lateral deviation of particles is significantly increased by crossing more parallel pressure nodal lines lying across the flow. In this system, the particles exhibit two distinct behaviors depending on the drag to acoustic radiation force ratio: a drift mode at low acoustic power and a locked mode at high acoustic power. Considering two types of particles, the maximum sorting efficiency is reached when one type of the particles is in the locked mode while the other is in the drift mode [4, 38, 40, 41, 43, 44]. Even though the transition from drift to locked mode has been observed numerically and experimentally, the transition threshold is not known in advance and there is still no theory to describe the trajectory of particles in drift mode. This absence of the theoretical model is a major hurdle for the design of high-throughput applications that operate at low acoustic to drag force ratio for a maximum flow-rate. Therefore, an accurate physical model to extract the key features of the particles in drift mode is highly desirable.

In this paper, we will summarize the key features of drift and locked modes, obtain an explicit transition threshold, and derive a model to predict analytically the trajectory of particles in each of them. Assuming that the microchannel has a large aspect ratio, the transition threshold and particle trajectories in locked mode are obtained using a nonlinear ordinary differential equation without further assumptions, while trajectories in the drift mode are derived using a perturbation analysis on the acoustic to drag force ratio. Analytical and numerical calculations are then showed in well agreement with previously published experimental data [25, 38]. Eventually, this analytical model proposes a reliable approach to evaluate the behavior of particles, and optimize the design of high-accuracy, high-sensitivity or high-throughput sorting microsystems.

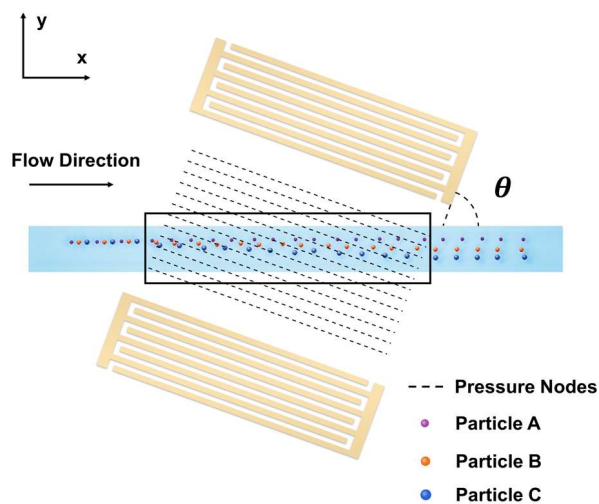

Figure 1. Schematic illustration of the acoustophoretic system. A pair of tilted-angle interdigitated electrodes generates propagative and counter-propagative surface acoustic waves to form an SSAW field in the channel. The vibrations generate a radiation pressure that deviates the particles or bio-samples in the flow. The propagative surface acoustic wave makes an angle $\theta$ with the flow. Three kinds of particles are presented in the schematic, where Particle A and C get the minimum and maximum lateral migrations, respectively.

## 2. Theory

### 2.1 Model derivation

At a few exceptions [42], the tilted-angle standing acoustic waves are generated by placing the sorting microchannel between a pair of surface acoustic wave transducers. The counter-propagating waves would generate a standing surface acoustic wave (SSAW) that spontaneously radiate in the fluid. We further assume that the particles only experience the standing-component of the acoustic field and are held in the mid-plane of the channel. This holding force might contribute to a vertical standing wave between the channel floor and ceiling (due to the reflections at the interface) [45], or nonlinear hydrodynamic focusing [46, 47].

For structures shown as Fig. 1, a particle immersed in an SSAW pressure field of negligible attenuation would experience an acoustic radiation force $\vec{F_{ac}}$ along the propagation direction of the acoustic wave

$$\vec{F_{ac}} = -A \sin(2\vec{k_{saw}} \cdot \vec{r_p}) \vec{k_{saw}} \tag{1}$$

where $\vec{k_{saw}}$ is the wave-vector from one of the transducer, $\vec{r_p}$ is the particle position vector, and $A$ is the magnitude of the radiation force (given in the appendix). A similar expression could also be obtained for any other force deriving from a scalar potential, including electrophoresis [19-21], dielectrophoresis [38] and optical trapping [13]. The factor $A$ may be positive or negative depending on the Clausius-Mosotti factor for dielectrophoresis, refraction index ratio in optophoresis, or the acoustic contrast factor for acoustophoresis.

The drag force $\vec{F_d}$ of viscosity on a solid spherical object, which moves through the fluid in a region relatively far from the channel walls, is given by the Stokes's law for conditions of small Reynolds number as [44]

$$\vec{F_d} = 6\pi\mu R(\vec{v_f} - \vec{v_p}) \tag{2}$$

where $\mu$ is the dynamic viscosity of the fluid, $R$ is the radius of the particle, and $\vec{v_f}$ and $\vec{v_p}$ are the velocity vectors of the medium fluid and the particle, respectively. Even though $\vec{v_f}$ is allowed to vary across the whole channel geometry, the calculations could be considerably simplified when it is constant. This condition is met in commonly used microchannels of high aspect ratio, where the boundary layer thickness (given by $h/\pi$ with $h$ the channel height) is much smaller than the channel width [48].

In the over-damped regime where all inertial effects are neglected, the forces applied on the particle are balanced in all directions, which yields $\vec{F_{ac}} + \vec{F_d} = \vec{0}$, as

$$-A \sin(2\vec{k_{saw}} \cdot \vec{r_p}) \cos\theta + 6\pi\mu R(v_f - v_{px}) = 0 \tag{3}$$

$$-A \sin(2\vec{k_{saw}} \cdot \vec{r_p}) \sin\theta - 6\pi\mu R v_{py} = 0 \tag{4}$$

where $v_{px}$ and $v_{py}$ are the magnitudes of the particle's speed along (longitudinal) and perpendicular (transversal) to the channel, respectively, and $\theta$ marks the angle between the channel and the wave-vector.

Assuming that the transversal position of particles ($y_p$) is only a function of its longitudinal position ($x_p$), the transversal velocity of particles could be expressed as

$$v_{py} = \frac{dy_p}{dt} = \frac{dy_p}{dx_p}\frac{dx_p}{dt} = y'_p v_{px} \tag{5}$$

Introducing the dimensionless radio of acoustic to drag force $\varepsilon = A/(6\pi\mu R v_f)$, Eqn 5 leads Eqn 3 and Eqn 4 to

$$[1 - \varepsilon \sin(2\overrightarrow{k_{saw}} \cdot \overrightarrow{r_p})\cos\theta]y'_p = -\varepsilon \sin(2\overrightarrow{k_{saw}} \cdot \overrightarrow{r_p})\sin\theta \tag{6}$$

This equation is first integrated numerically with the BDF solver of scipy.integrate.solve_ivp, setting $k = 1$, $\theta = 75°$, and $\varepsilon$ ranging from $-0.1$ up to $0.26$ ($\approx \cos\theta$). The resulting trajectories are shown in Fig. 2. When the acoustic to drag force ratio is small ($\varepsilon = \pm 0.1$ in this example), the particles essentially follow the flow stream while weakly oscillating as they cross the ridges and valleys of the acoustic potential landscape. However, closer observation reveals that the particles would also slowly drift towards the negative y-direction: this is the drift mode described earlier. We also note that even though an opposite sign of $\varepsilon$ (that might be due to opposite acoustic contrast factor for instance) results in anti-phase oscillations, it yields the same drift velocity and a mirrored path. Next, at higher acoustic to drag force ratio ($\varepsilon = 0.2$), the oscillation becomes sharper and the particle's drift in the acoustic field is more obvious. Eventually, once the acoustic force exceeds the drag force, the particles will stall against the acoustic ridges and therefore be locked in the valleys. In this locked-mode ($|\varepsilon| \geq 0.26$ in this example), the particles would move parallel to the acoustic wavefronts.

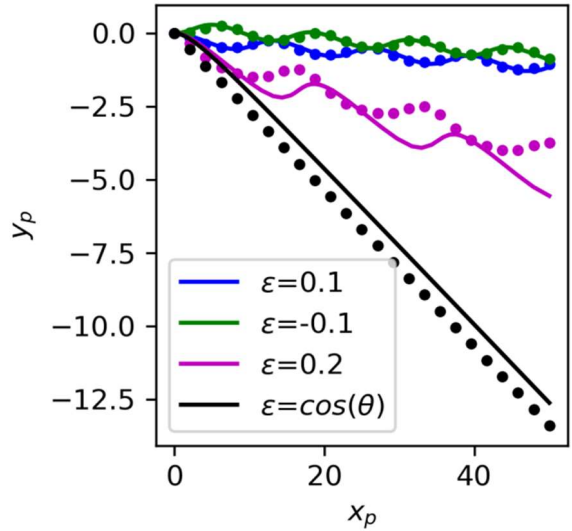

Figure 2. Comparison of analytical and numerical results for the trajectory at different acoustic to drag force ratio. The parameters are $k = 1$ and $\theta = 75°$ ($\cos\theta \approx 0.26$). The numerical solutions of Eqn. 6 are given by solid lines, and the analytical solutions are given by the regularly spaced dots. $|\varepsilon| = |\cos\theta|$ outlines the transition threshold between drift mode and locked mode. Particles with opposite contrast factors tend to follow opposite oscillation patterns in drift mode (shown by $\varepsilon = \pm 0.1$). The transition from drift to locked mode ($\varepsilon = 0.2$) is strongly non-linear and is not studied here.

## 2.2 Locked-mode trajectory

As first pointed out by the pioneering work of Collins et al. [38], particles exposed to a tilted-angle SSAW normally exhibit two distinct behavior depending on the value of $\varepsilon$ (defined as locked mode and drift mode as shown in Fig. 2), while the transition threshold between these two modes was estimated by $\varepsilon \geq \cos^4 \theta$. Starting from the assumption that the particles move in straight lines in the locked mode, we obtain a different value for the transition threshold (derivation in the appendix):

$$|\varepsilon| \geq |\cos \theta| \tag{7}$$

The trigonometric calculation also presents the straight path for particles in locked-mode:

$$y_p = y_0 - \frac{1}{\tan \theta} x_p \tag{8}$$

with $y_0$ the original transversal position of the particle in the channel. We note a good agreement between Eqn. 8 and the numerical solution of Eqn. 6 as shown in Fig. 2. The small shift between them is attributed to the $\sin(\psi_\infty)$ term defined in the appendix, which is neglected here for the sake of simplicity.

## 2.3 Drift-mode trajectory

While the locked-mode is fairly well understood, the drift mode has long been overlooked due to its nonlinearity and complexity, despite its practical importance in high-throughput applications [25, 37, 39-43]. Even though Eqn. 6 can be solved numerically, an analytical study can reveal the main features more clearly and yields simple estimates of the trajectory.

For particles crossing parallel pressure nodal lines in the drift mode, they are exposed to an oscillating acoustic radiation force. Considering that a linear expansion would only yield an oscillating motion with a zero average, the slow drift of particles is a cumulative effect that can only be studied by, at least, second-order terms. Hence, we expand Eqn. 6 up to the second-order expression of $\varepsilon$:

$$y_p' = -\varepsilon \sin(2\vec{k_{saw}} \cdot \vec{r_p}) \sin \theta \left[1 + \varepsilon \sin(2\vec{k_{saw}} \cdot \vec{r_p}) \cos \theta\right] \tag{9}$$

To solve this equation, a candidate solution is proposed as

$$y_p = y_0 + \alpha_0 x_p + \varepsilon(y_1 + \alpha_1 x_p) + \varepsilon^2(y_2 + \alpha_2 x_p) \tag{10}$$

where $y_{0,1,2}$ are periodic functions of $x_p$, and $\alpha_{0,1,2}$ are constants to represent the drift. After substitution of Eqn. 10 in Eqn. 9, we could search for non-trivial solutions by grouping the contributions of equal powers in $\varepsilon$ (the process is outlined in the appendix). That yields:

$$y_0' + \alpha_0 = 0 \tag{11}$$

$$y_1' + \alpha_1 = -\sin(\Omega) \sin \theta \tag{12}$$

$$y_2' + \alpha_2 = -\sin^2(\Omega) \sin \theta \cos \theta - 2k_y y_1 \cos(\Omega) \sin \theta \tag{13}$$

with $\Omega = 2k_x x_p + 2k_y(y_0 + \varepsilon^2 \alpha_2 x_p)$. Equations (11-13) are solved iteratively then as described in the appendix, which gives

$$y_p = y_0 + \frac{\varepsilon}{2k_x}\cos(\Omega)\sin\theta + \varepsilon^2 y_2 + \varepsilon^2 \alpha_2 x_p \quad (14)$$

with $\alpha_2 = -\tan\theta / 2$ and $y_2$ provided in the appendix. It should be noted that $\varepsilon^2 y_2$ is very small and may be neglected for rapid evaluations of the deviation.

Eqn. 14 is composed of 3 terms of decreasing magnitude: the original transversal position of the particle $y_0$ is the only non-vanishing term at zero acoustic power, $[\varepsilon \cos(\Omega)\sin\theta] / (2k_x)$ represents the oscillations of the particle as it crosses the washboard-like acoustic potential, and $\varepsilon^2 \alpha_2 x_p$ accounts for the cumulative lateral migration. We note that even though the oscillating magnitude depends on $\varepsilon$ (i.e. opposite acoustic contrast factor will lead to an opposite oscillation pattern), the particles will always migrate in the same direction regardless of the sign of the acoustic contrast factor. The drifting slope should be proportional to the square of the acoustic power, the sixth power of the particle radius, and reciprocal of the square of the flow velocity. Even though this provides a very sharp cutoff for the particle size which is comparable to travelling wave systems, it also means that the sorting performance is very sensitive to the flow rate. Besides, $\alpha_2$ is periodic of period $\pi$, meaning that it is insensitive to the arbitrary choice of which transducer generates the traveling wave of wave-vector $\overrightarrow{k_{saw}}$. Since $d\alpha_2/d\theta = -1 / (2\cos^2\theta)$, we deduce that the efficiency of drift mode diverges to infinity when $\theta = (2n+1)\pi/2$ for any integer $n$ (particles are more easily deflected when the wavefronts are parallel to the flow). Finally, since $|\varepsilon| \geq |\cos\theta|$ marks the transition to locked mode, we have $|\varepsilon^2 \alpha_2| < |0.25 \sin 2\theta|$. Therefore, the highest slope for the drift mode is obtained at $\theta = \pm 45°$. Indeed, at smaller $\theta$ the particles exhibit smaller oscillations while at higher $\theta$ they tend to switch to the locked mode for smaller values of $\varepsilon$.

Finally, we could get the deviation slope in drift and locked modes:

$$S = S_{\text{drift}} = -\varepsilon^2 \alpha_2, \quad \text{if } |\varepsilon| < |\cos\theta| \quad (15)$$

$$S = S_{\text{locked}} = -\frac{1}{\tan\theta}, \quad \text{if } |\varepsilon| \geq |\cos\theta| \quad (16)$$

3. **Model accuracy evaluation**

Since $k$ is only a scaling factor in Eqn. 6, $\theta$ and $\varepsilon$ are the only relevant parameters to predict the trajectory. In Fig. 3, the decimal logarithm of the slope predicted by the analytical model is compared with the numerical calculations of the Eqn. 6. The difference in the drift region is often negligible except when $\varepsilon$ approaches $\cos\theta$. $\varepsilon = \cos\theta$ is denoted as the dotted line, beyond which the particle will be captured and migrate along the acoustic pressure valleys. The predicted trajectories obtained by our analytical model behave in a good agreement with the described numerical solution in the considered range of parameters.

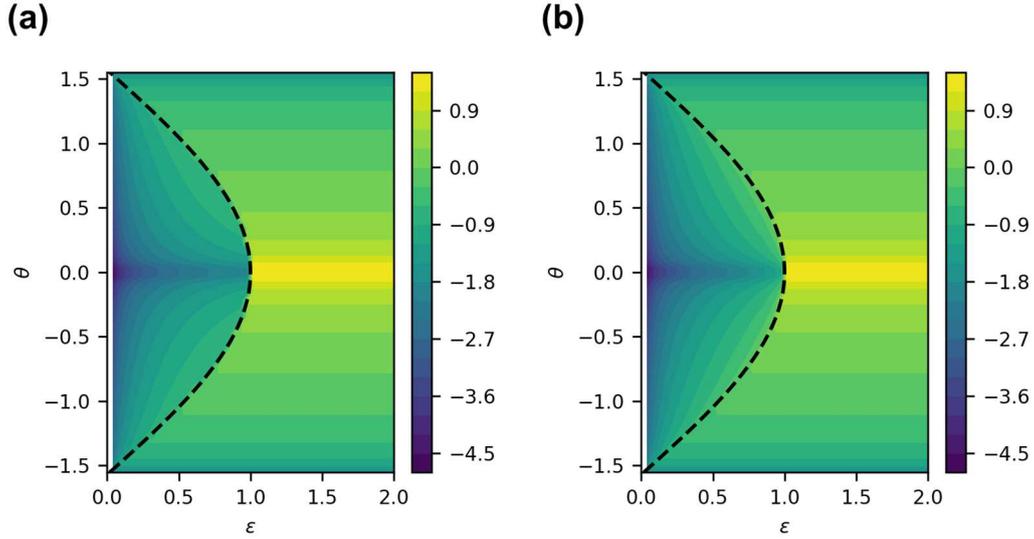

Figure 3. Comparison of analytical (left) and numerical (right) estimation of the deviation slope. The value of $\varepsilon$ and $\theta$ are presented by the *x*- and *y*-axis of the coordinate, respectively. The color scale is given by $\log_{10} S$. The dotted line ($\varepsilon = \cos\theta$) marks the transition threshold from drift to locked mode.

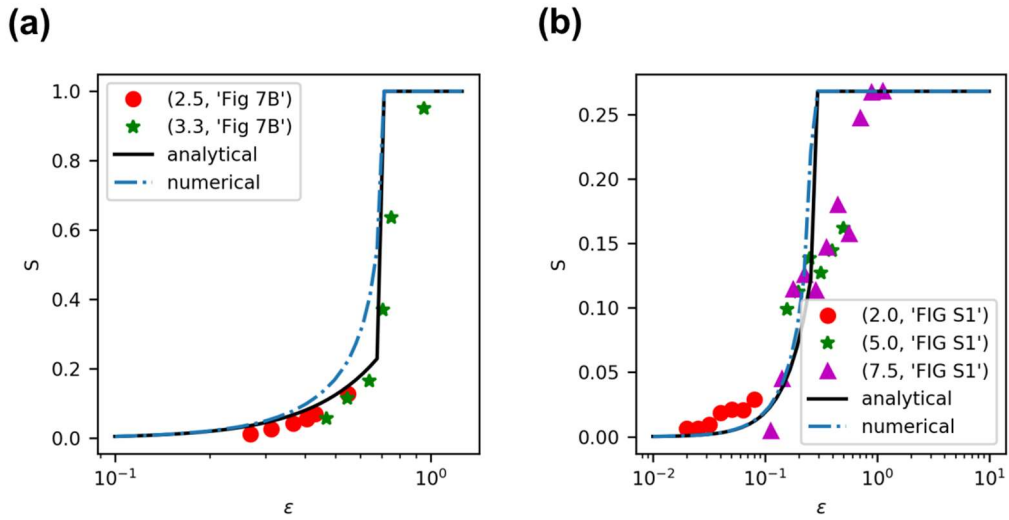

Figure 4. Comparison of the model with experimental results from (a) Collins et al. [38] and (b) Ding et al. [25] while adopting their structural and operating setups. $\varepsilon$ scales the *x*-axis in the coordinate while *y*-axis demonstrates the slope of the trajectory. The experimental, analytical, and numerical solutions are presented by symbols, solid lines, and dotted lines, respectively.

As the analytical results have been validated against numerical integration, we now compare it to experimental data derived from two influential papers [25, 38]. The value of $\varepsilon$ is calculated from factor $A$ (given in the appendix). We derive the acoustic energy density by assuming that it is proportional to the electrical power reported by the authors, and the proportionality coefficient $\beta$ is constant for each microfluidic device (in agreement with the authors' guidelines). Since the analyzed data in each paper comes from a single microfluidic device, only two different values of $\beta$ have to be determined. Each constant is obtained by fitting the experimental particles slope to our analytical model with a brute-force algorithm. Fig. 4 compares experimental, numerical (calculated from Eqn. 6) and analytical (calculated from Eqn. 15 and Eqn. 16) estimates of the deviation slope. Generally speaking, our theory is in a satisfactory agreement with their experimental results within the assumed bounds (i.e. $\varepsilon \ll 1$ and $|\varepsilon| < |\cos\theta|$) and a moderate deviation in the transition regime. This is in agreement with the finding of previous authors [25, 38] that Eqn. 6 reproduces well the experimental data.

## 4. Optimal design and operation parameters

Having validated our calculations against experimental data, we may now analyze the optimal choice of structural (the tilted-angle) and operating parameters (the acoustic power) to maximize the sorting efficiency for two populations of particles. Taking the acoustic to drag force ratio for the second kind of particle as $\varepsilon_2 = m\varepsilon$, we assume $m < 1$ (the types of particles could be swapped to fulfill this condition). The locked mode offers the maximum deviation for one type of particles, while the deviation of the second type of particles could then be given by $\varepsilon_2^2 \alpha_2 = 0.25 m^2 \sin 2\theta$, so the deviation slope difference $\Delta S$ reads:

$$\Delta S = \frac{1}{\tan\theta} - m^2 \frac{\sin 2\theta}{4} \tag{17}$$

Hence, the highest sorting efficiency is obtained with $\theta = 0$, that is when the wavefronts are orthogonal to the flow. This also requires the highest ratio of acoustic to drag force ($\varepsilon = 1$) which may not always be possible in practice due to thermal management constraints. More generally, $\Delta S$ is a monotonically decreasing function of $\theta$ meaning that $\theta$ should be set as high as possible as long as $|\varepsilon| = |\cos\theta|$ can be satisfied.

In some applications, such as high-throughput devices, the acoustic to drag ratio may be very small. According to Eqn. 7, it is always possible to let $\theta$ approach 90° such that the particles could travel in locked mode. The major shortcoming of this strategy is that the difference between drift and locked-mode trajectories is also decreasing to 0. Expanding $\Delta S$ up to the first order in $\pi/2 - \theta$, we could get:

$$\Delta S \underset{\theta \to \frac{\pi}{2}}{\approx} \left(\frac{\pi}{2} - \theta\right)\left(1 - \frac{m^2}{2}\right) \tag{18}$$

meaning that drift and locked mode deviation decrease at the same rate when $\theta$ approaches $90^o$.

Eventually, since the particles deviation in drift mode is highly sensitive to the particle size and the acoustic contrast factor, monitoring the deviation of particles in drift mode could potentially have sensing applications. For such applications, the highest possible deviation in drift mode is obtained at $\theta = 45^o$. All these considerations are summarized in table 1:

Table 1. Performance of the separation system for different tilted angles.

| $\theta$ | 0 | $45^o$ | $90^o$ |
|---|---|---|---|
| $\varepsilon_c = |\cos\theta|$ | 1 | $1/\sqrt{2}$ | 0 |

| | | | |
|---|---|---|---|
| $\max(|S_{drift}|)$ | 0 | 1/4 | 0 |
| $\max(|S_{lock}|)$ | ∞ | 1 | 0 |
| **Feature** | Largest difference of slope between drift and locked mode | Largest possible slope for the drift mode | Lowest drift-to-lock transition threshold |
| **Proposed application** | High precision sorting (requires high power or low throughput) | Maximum sensitivity (measurement of $\varepsilon$) | High-throughput sorting or low-power sorting (limited precision) |

## 5. Conclusion

Particle sorting with the tilted-angle standing surface acoustic wave is a promising method for the enrichment, purification or extraction of bio-samples and particles in microfluidic systems. In this paper, a thorough theoretical study has been conducted. Two regimes of particle deviation have been identified as the drift mode and the locked mode that occur at low and high acoustic to drag force ratio, respectively. The trajectory of particles in each mode is deduced analytically, with an explicit transition threshold. The proposed model reveals that deviation in the drift mode is fundamentally nonlinear, whose slope is proportional to the sixth power of the particle radius, the fourth power of the acoustic pressure, the square of the acoustic contrast factor, and reciprocal of the square of the flow velocity. Our analytical formula showed a good agreement with numerical calculations and experimental data even in the transition regime. The model provides a reliable approach to evaluate the choice of structural and operating parameters for the best performance of the separation system.

According to the derived model, the maximum sorting precision is obtained when the acoustic wavefronts are orthogonal to the flow, but this also requires the highest level of acoustic power. Since the deviation in drift mode is highly sensitive to the particle size and acoustic contrast, we envision that the intermediate situation of $45^o$ tilt acoustophoresis devices yields the widest possible span of drift-mode deviation, and could be used for measuring particle size and acoustic contrast.

**Acknowledgements**

This work was supported by the National Science Foundation of China with Grant No. 61704169, the National Natural Science Foundation of China with Grant No. 61874033, the Natural Science Foundation of Shanghai Municipal with Grant No. 18ZR1402600, and the State Key Lab of ASIC and System, Fudan University with Grant No. 2018MS003. The work was also supported by the China Scholarship Council (CSC) with File No. 201706100083.

# Appendix

## A1. Value of A

When particles much smaller than the acoustic wavelength are exposed to a plane standing acoustic waves or to the radiation issued from plane standing surface acoustic waves, the force magnitude reads [44, 49]:

$$A = \frac{4\pi}{3} R^3 E_{ac} \varphi, \quad (A1)$$

$$E_{ac} = \frac{p_0^2}{4\rho_0 c_0^2}, \quad (A2)$$

where $E_{ac}$ is the acoustic energy density and $\varphi$ is the acoustic contrast factor of the particle. $R$ denotes the radius of the particle. $c_0$ and $\rho_0$ are the density and acoustic velocity in the fluid medium. The acoustic contrast factor for the surface acoustic wave differs from the usual expression with bulk acoustic waves:

$$\varphi = f_1^r + \frac{3}{2} f_2^r \frac{k_s^2 - k_z^2}{k^2} \quad (A3)$$

$$p_0^2 = \frac{\beta P_e \rho_s c_{SAW} k_{SAW}}{A_{IDT}} \quad (A4)$$

The wavenumber in water $k$ is given by $k^2 = \omega^2/c_0^2$. Using the dispersion relation, one obtains the vertical wavenumber $k_z^2 = k^2 - k_s^2$ with $k_s$ the wavenumber. For acoustic waves along the lithium niobate/water interface, the scaling factor $(k_s^2 - k_z^2)/k^2$ is approximately 0.704 [44]. The terms $f_1^r$ and $f_2^r$ refer to the real part of the monopolar and dipolar scattering coefficients $f_1$ and $f_2$, respectively. Given the size of the target particles is much greater than the thickness of the viscous and thermal boundary layers, we have $f_1^r = 1 - \tilde{\kappa}$ and $f_2^r \approx 2(\tilde{\rho} - 1)/(2\tilde{\rho} + 1)$ with $\tilde{\kappa}$ and $\tilde{\rho}$ the compressibility and density ratios of the particle and the fluid, respectively [31]. The acoustic pressure $p_0$ was evaluated as $p_0^2 = \beta P_{IN} \rho_s c_{SAW} k_{SAW} / A_{IDT}$, where $P_e$ is the power of the IDTs, $\rho_s$ and $c_{SAW}$ are the density and the sound speed of the substrate, respectively, $A_{IDT}$ is the aperture of IDTs multiplied by their distance, and $\alpha$ is the power conversion efficiency in which the power of the IDTs converts to the acoustic pressure in the fluids.

## A2. Locked-mode threshold and solution

The locked mode is defined as the power level when the particles cannot overcome the acoustic ridge and jump from node to node. They are therefore expected to move along straight lines. We propose a candidate solution at high power:

$$y_p = y_0 + \alpha_\infty x_p \quad (A5)$$

where $\alpha_\infty$ has to be determined. Using Eqn. 6 we get:

$$[1 - \varepsilon \sin(2\overrightarrow{k_{saw}} \cdot \overrightarrow{r_p}) \cos \theta] \alpha_\infty = -\varepsilon \sin(2\overrightarrow{k_{saw}} \cdot \overrightarrow{r_p}) \sin \theta \qquad (A6)$$

This equation can only be satisfied when the phase $2\overrightarrow{k_{saw}} \cdot \overrightarrow{r_p}$ is a constant $\psi_\infty$. Therefore we get $2k_x x_p + 2k_y \alpha_\infty x_p + 2k_y y_0 = \psi_\infty$ whatever the value of $x_p$, thus $\alpha_\infty = -k_x/k_y = -1/\tan(\theta)$. Then Eqn. A6 becomes:

$$\frac{1 - \varepsilon \sin(\psi_\infty) \cos \theta}{\tan(\theta)} = \varepsilon \sin(\psi_\infty) \sin \theta \qquad (A7)$$

which, after a series of elementary trigonometric manipulations, yields:

$$\varepsilon \sin(\psi_\infty) = \cos(\theta) \qquad (A8)$$

Depending on the acoustic contrast, $\varepsilon$ may be either positive or negative, therefore one has to consider $|\varepsilon|$. Provided that $|\varepsilon| \geq |\cos \theta|$, it is possible to find a $\varphi_\infty$ to satisfy Eqn. A8, so particles will follow a straight line beyond this limit. Hence, the threshold for the transition to the locked mode is given by Eqn. 7.

### A3. Drift mode solution

The special choice of candidate solution for the drift mode is advantageous for two reasons. First, it can accommodate large transversal migration: even though the drift parameter $\varepsilon$ may be very small, the channel may be very long (that is $x_p$ can be large) and therefore the cumulative drift given by $\varepsilon \alpha_1 x_p$ and $\varepsilon^2 \alpha_2 x_p$ are allowed to be quite large. In addition, for any periodic function $y$ of period $L$:

$$\frac{1}{L} \int_0^L y'(x) \, dx = y(L) - y(0) = 0 \qquad (A9)$$

This gives a convenient way to isolate periodic and non-periodic components in the resolution of the equations. However, the phase $2\overrightarrow{k_{saw}} \cdot \overrightarrow{r_p} = 2k_x x_p + 2k_y y_p = 2k_x x_p + 2k_y [y_0 + \alpha_0 x_p + \varepsilon(y_1 + \alpha_1 x_p) + \varepsilon^2(y_2 + \alpha_2 x_p)]$ must be discussed more extensively: even though the drifting regime ensures that $\varepsilon \ll 1$, the channel may be very long so that $2\varepsilon \alpha_1 k_y x_p$ and $2\varepsilon^2 \alpha_2 k_y x_p$ cannot be neglected. Simple trigonometry allows segregating smaller and larger contributions:

$$\sin(2\overrightarrow{k_{saw}} \cdot \overrightarrow{r_p}) = \sin(\Omega) \cos(\varepsilon y_1 + \varepsilon^2 y_2) + \cos(\Omega) \sin(\varepsilon 2k_y y_1 + \varepsilon^2 2k_y y_2) \qquad (A10)$$

with $\Omega = 2k_x x_p + 2k_y[y_0 + \alpha_0 x_p + \varepsilon \alpha_1 x_p + \varepsilon^2 \alpha_2 x_p]$.

During the calculations, we will only need the first order in $\varepsilon$:

$$\sin(2\overrightarrow{k_{saw}} \cdot \overrightarrow{r_p}) = \sin(\Omega) + \varepsilon 2k_y y_1 \cos(\Omega) \qquad (A11)$$

Taking the average of Eqn. 13 and using Eqn. 12, we find that $\alpha_0 = 0$ and $y_0$ is a constant giving the initial transversal position of the particle. Similarly to the calculation of $\alpha_0$, we get $\alpha_1 = 0$. Then, $y_1$ is integrated as:

$$y_1 = \frac{1}{2k_x}[\cos(\Omega) - \cos(\Omega_0)]\sin\theta \quad (A12)$$

where $\Omega_0 = \Omega(x_p = 0)$ and the $2k_y\varepsilon^2\alpha_2$ term in the denominator has been neglected (because the $y_1$ is already zero order of $\varepsilon$).

Similarly to the calculation of $\alpha_0$, we obtain after average $\alpha_2 = -\tan\theta/2$, while $y_2$ reads:

$$y_2 = -\frac{1}{8k_x}[\cos(2\Omega) - \cos(2\Omega_0)]\sin\theta\,[\cos\theta + \sin\theta\tan\theta] \quad (A13)$$